\documentclass[a4paper,12pt]{article}
\usepackage{amsmath}
\usepackage{amssymb}
\usepackage{accents}
\usepackage{tikz}
\newtheorem{thm}{Theorem}[section]
\newtheorem{prop}[thm]{Proposition}

\newtheorem{definition}[thm]{Definition}

\def\C{{\mathbb C}}
\def\P{{\mathbb P}}
\def\Z{{\mathbb Z}}

\def\dfrac#1#2{{\displaystyle\frac{#1}{#2}}}

\newcommand{\doublehat}[1]{\hat{\hat{#1}}}
\newcommand{\id}{\mathrm{id}}
\begin{document}

\begin{center}
{\LARGE Quadrirational Yang-Baxter maps \\[0.3em] and the $\tilde{E}_8$ Painlev\'e lattice}
\vskip10mm
{\large James Atkinson$^{1,2}$ and Yasuhiko Yamada$^3$}\\
\end{center}
\vskip5mm
{\small 
1. Department of Mathematics, Faculty of Engineering and Environment, Northumbria University, Newcastle upon Tyne, UK.\\
2. School of Mathematics and Statistics, the University of Sydney, NSW 2006, Australia.\\
3. Department of Mathematics, Faculty of Science, Kobe University, 657-8501, Japan.\\
}
\vskip5mm
\noindent
{\bf Abstract.}
We establish that the quadrirational Yang-Baxter maps, considered on their symmetry-complete lattice, give an un-normalized form of the Painlev\'e systems associated with $\tilde{E}_8$ symmetry.
This is a unified representation bringing KdV-type and Painlev\'e-type systems together outside of the usual paradigm of reductions. 
Our approach exploits the geometric characterisation of the Painlev\'e equations and the formulation of both kinds of systems in terms of birational groups.
\vskip5mm




\section{Introduction}\label{INTRO}
The quadrirational mappings \cite{ABSf} are a few amongst the myriad known manifestations of the discrete KdV equation (see \cite{PTV}).
However, three of them, denoted by $F_I$, $F_{II}$ and $F_{IV}$ in \cite{ABSf}, are set apart due to a seemingly coincidental feature: admissible transformations allow the system parameters to be put on the same footing as the variables.
This parameter-variable interchange symmetry is inessential for the KdV-type integrability, but turns out to signify existence of a larger whole of which the system is a part.
Specifically there is a more symmetric embedding lattice in which there is no distinction between parameters and variables on a global level \cite{ib}.
The KdV-type integrability is related to consistent embedding in the hypercube, which locally is the Yang-Baxter property, or consistency on a cube.
The corresponding local property fundamental for the extended lattice is a consistency which generalizes the cube combinatorics to that of the 5-simplex.
The extended lattice is characterised in general by its automorphism group, which is of Coxeter type.

Lattices with similar groups of automorphisms are well-known from the setting of the discrete Painlev\'e equations, and it is the purpose of the present paper to show the explicit connection.
Specifically, we will identify the lattice systems emerging from $F_I$, $F_{II}$ and $F_{IV}$ as being a representation of those systems from Sakai's Painlev\'e classification \cite{sc} with $\tilde{E}_8$ symmetry.
The key to making contact with the Painlev\'e systems is their manifestation as birational groups.
This formulation is basic to the geometric approach that has been developed by Sakai in the classification, it arises as the automorphism group of the underlying rational surface.
An advantage of the birational group is that it can be expressed in terms of simple generators, whilst encoding the Painlev\'e dynamics and B\"acklund transformations on an equal footing.

Section \ref{5SC} summarises relevant theory of the 5-simplex consistent systems, in particular the associated birational actions are defined, as well as the associated lattice and initial-value-problem. 
We also explain how the better-known setting for the Yang-Baxter maps is recovered by restriction to a sub-lattice.
Section \ref{LTI} starts with the main technical result, which is a substitution to linearize the previously defined birational actions in the case corresponding to $F_I$.
This gives direct connection with the $\P^1\times \P^1$ actions associated with the elliptic Painlev\'e equation, which arise here in a form most similar to those defined in \cite{nty} (see also \cite{msy,sc,kmnoy}).
We then find that the cases $F_{II}$ and $F_{IV}$ can be treated similarly, in the Painlev\'e setting they correspond to different parameterisations of the elliptic curve.
We finish with some remarks about the lattice-geometry.

\section{The 5-simplex consistent systems}\label{5SC}
The purpose of this section is to summarise the relevant theory of the 5-simplex consistent systems, which are a subset of the quadrirational Yang-Baxter maps having additional symmetry.
More details can be found in \cite{ABSf,ib}.
\subsection{5-simplex consistency}
The basic object is the rational triplet-pair system.
\begin{definition} \label{tps}
By a rational triplet-pair system we mean a system of equations relating a pair of triplets of variables \[x_0,x_1,x_2,\quad y_0,y_1,y_2,\] that rationally determines one variable from each triplet from the remaining four variables, and which is invariant under permutations of the variables that send the set $\{\{x_0,x_1,x_2\},\{y_0,y_1,y_2\}\}$ to itself.
\end{definition}
The explicit examples may be written as follows:
\begin{eqnarray}
&(1-x_0)(1-x_1)(1-x_2) = (1-y_0)(1-y_1)(1-y_2),\quad x_0x_1x_2=y_0y_1y_2,\ \label{F1a}\\
&x_0x_1x_2=y_0y_1y_2, \quad x_0+x_1+x_2=y_0+y_1+y_2,\ \label{F2a}\\
&x_0+x_1+x_2=y_0+y_1+y_2,\quad x_0^2+x_1^2+x_2^2=y_0^2+y_1^2+y_2^2.\ \label{F4a}
\end{eqnarray}
From these expressions the triplet-pair symmetry is clear, and the rationality is straightforward to verify.
With one variable from each triplet (e.g. $x_0$ and $y_0$) playing a distinguished passive role as parameters, these systems are equivalent to the quadrirational Yang-Baxter maps $F_I$, $F_{II}$ and $F_{IV}$ from \cite{ABSf}.
Based on that classification (of quadrirational systems), it is likely that these examples exhaust the scalar systems satisfying Definition \ref{tps}.

For these cases, the parameter-dependent Yang-Baxter property turns out to be a subcase of a stronger consistency feature in which the parameters are elevated to the same level as variables.
The combinatorics of this consistency can be described in terms of the 5-simplex.
The 5-simplex is a geometric figure consisting of six vertices in which every pair of vertices defines a single edge.
It naturally occupies a space of 5 dimensions, but only the combinatorics are relevant here, and not the spatial geometry.
\begin{definition}\label{5sc}
Assign variables to edges of the 5-simplex and impose a rational triplet-pair system (Definition \ref{tps}) on each pair of triplets of variables coming from the pairs of vertex-disjoint triangular faces.
There are 6 edges along a closed path that passes once through each vertex of the 5-simplex, if the variables coming from those edges are unconstrained by the imposed equations, then the triplet-pair system is called 5-simplex consistent.
\end{definition}
The 5-simplex has 15 edges and 10 vertex-disjoint triangle-pairs.
It is easy to check that a generic rational triplet-pair system allows to determine all remaining edge variables in terms of the subset of 6 variables along the closed path.
In the generic case the variables will be overdetermined because there are 9 remaining, and 20 imposed equations (coming from the 10 triangle-pairs), however:
\begin{prop}
The rational triplet-pair systems (\ref{F1a}), (\ref{F2a}) and (\ref{F4a}) are 5-simplex consistent.
\end{prop}
The relevant structure resulting from this consistency property can be viewed as a birational group, or an equivalent lattice system.
The two points of view are related by a two-way correspondence, connecting elements of the birational group with combinatorial automorphisms of the lattice.
It is convenient to begin by introducing the birational group initially, and the lattice system afterwords.
\subsection{Associated birational group}\label{ABG}
Given an arbitrary positive integer $n$ and a generic rational triplet-pair system (Definition \ref{tps}), the associated birational group is defined as follows.
Fix $j\in\{1,\ldots,n\}$, impose the rational triplet-pair system on all triplet-pairs
\begin{equation}
y_j,x_i,\bar{x}_i, \quad x_j,y_i,\bar{y}_i, \qquad i \in \{1,\ldots,n\}\setminus\{j\},\label{mc}
\end{equation}
and set $\bar{x}_j=x_j,\bar{y}_j=y_j$.
This system of equations determines the variables 
\begin{equation}\label{Fvars}
\bar{x}_1,\ldots,\bar{x}_n,\bar{y}_1,\ldots,\bar{y}_n
\end{equation}
as rational functions of the variables 
\begin{equation}\label{Evars}
{x}_1,\ldots,{x}_n,{y}_1,\ldots,{y}_n.
\end{equation}
A rational mapping $\sigma_j$, acting on the $2n$ variables (\ref{Evars}), is introduced as
\begin{equation}\label{sigmadef}
\sigma_j:= \{x_i\mapsto \bar{x}_i, \ i\in\{1,\ldots,n\}\setminus \{j\} \}.
\end{equation}
We adopt the convention of omitting trivial actions, thus $\sigma_j$ leaves unaltered the variables $x_j$ and $y_1,\ldots,y_n$.
Denoting the permutation mapping which interchanges $x$ and $y$ variables by $\omega$,
\begin{equation}\label{omegadef}
\omega:= \{x_i\leftrightarrow y_i, \ i\in\{1,\ldots,n\} \}
\end{equation}
allows, by conjugation, to introduce the companion to $\sigma_j$.
We denote it by $\sigma_j^\omega:=\omega\sigma_j\omega^{-1}$:
\begin{equation}
\sigma_j^\omega= \{y_i\mapsto \bar{y}_i, \ i\in\{1,\ldots,n\}\setminus \{j\} \}.
\end{equation}
The following relations satisfied by the mappings are clear either from the definitions, or from the symmetry of the rational triplet-pair system:
\begin{equation}\label{rels1}
\sigma_i^2=\omega^2=(\sigma_i\omega)^4=\id.
\end{equation}
According to our notational convention, $\id:=\{\}$, it is the identity mapping.

On the level of these constructed mappings, the 5-simplex consistency manifests as further group relations.
\begin{prop}
Let $m$ and $n$ be integers such that $2<m<n$.
The underlying rational triplet-pair system is 5-simplex consistent (Definition \ref{5sc}) if and only if the derived mappings $\sigma_1,\ldots,\sigma_m$ (\ref{sigmadef}) and mapping $\omega$ (\ref{omegadef}) satisfy the additional relations
\begin{equation}
\begin{split}
(\sigma_i\sigma_j)^2 = (\sigma_i\omega\sigma_j\omega)^3 = \id, & \qquad |\{i,j\}|=2,\\
((\sigma_i\sigma_j\omega)^2\sigma_k\omega)^2 = \id, & \qquad |\{i,j,k\}|=3.
\end{split}
\label{rels2}
\end{equation}
\end{prop}
These relations define a Coxeter group.
\begin{prop}\label{vmg}
Suppose $m>1$ and consider the finitely presented group with generators $\sigma_1,\ldots,\sigma_m,\omega$ and relations (\ref{rels1}), (\ref{rels2}).
Elements $t_0,\ldots,t_{m+2}$ defined as
\begin{equation}
\begin{split}
t_0 & = \sigma_1,\\
t_1 & = \omega,\\
t_2 & = \sigma_1\omega\sigma_1,\\
t_3 & = \omega\sigma_1\sigma_2\omega\sigma_2\sigma_1\omega,\\
t_{i+3} & = (\omega\sigma_i\sigma_{i+1})^3, \qquad i\in\{1,\ldots,m-1\},
\end{split}\label{iso}
\end{equation}
are alternative generators for the group. 
In terms of these generators, the relations are encoded in the diagram 
\[
\begin{array}{cccccccccccccc}
&&&& t_2\\
&&&& \vert\\
&&&& t_4\\
&&&& \vert\\
t_1&-&t_3&-&t_5&-&t_6&-&\cdots&-&t_{m+2}\\
\end{array}
\]
together with diagram automorphism through the action of $t_0$,
\begin{equation}
\begin{split}
t_0^2 & = \id,\\
t_0 t_1 & = t_2 t_0,\\
t_0 t_3 & = t_4 t_0,\\
t_0 t_i & = t_i t_0, \qquad i \in \{5,\ldots,m+2\}.\\
\end{split}\label{ecr}
\end{equation}
\end{prop}
{\bf Remark.} If the underlying rational triplet-pair system is 5-simplex consistent, then $t_4,\ldots,t_{m+2}$ act purely by permutation, $t_{i+3}=(\omega\sigma_i\sigma_{i+1})^3=\{x_i\leftrightarrow x_{i+1},y_i\leftrightarrow y_{i+1}\}$, $i\in\{1,\ldots,m-1\}$.
Thus the alternative Coxeter-type generators $t_0,\ldots,t_{m+2}$ are often simpler to deal with than $\sigma_1,\ldots,\sigma_m,\omega$, and they are natural for establishing contact with the Painlev\'e systems later.
On the other hand, there turns out to be a natural correspondence between the group elements $\sigma_1^\omega,\ldots,\sigma_m^\omega$, $\sigma_1,\ldots,\sigma_m$ and the variables (\ref{Evars}), and this is used to our advantage in the coordinatisation of the associated lattice in the following subsection.

\subsection{Associated lattice system}\label{AL}
The lattice for the rational triplet-pair systems is specified as a set of variables arranged into triplet-pairs.
A key feature of such arrangement is its group of symmetries, by which we mean the permutations of the variables which also permute the triplet-pairs.
For the lattice in question it corresponds to a natural permutation action of the group defined in Proposition \ref{vmg}.
\begin{definition}\label{Gtp}
For integer $m>1$, consider the finitely presented group defined by generators $\sigma_1,\ldots,\sigma_m,\omega$ and relations (\ref{rels1}), (\ref{rels2}) (cf. Proposition \ref{vmg}).
The associated triplet-pair arrangement is a set of variables assigned to elements of the conjugacy class of $\sigma_1$, which can be written as follows
\begin{equation}\label{vars}
w(\sigma_1^g), \quad g\in \langle t_1,\ldots,t_{m+2} \rangle,
\end{equation}
arranged into the triplet-pairs
\begin{equation}
w(\sigma_1^{g}),w(\sigma_2^{g\omega}),w(\sigma_2^{g\sigma_1\omega}), \quad
w(\sigma_2^{g}),w(\sigma_1^{g\omega}),w(\sigma_1^{g\sigma_2\omega}), \quad
g\in \langle t_1,\ldots,t_{m+2} \rangle.\label{gc}
\end{equation}
\end{definition}
This definition is consistent because $\sigma_1$ and $\sigma_2$ are conjugate.
Also it is clear form Proposition \ref{vmg} that $\langle \sigma_1,\ldots,\sigma_m,\omega \rangle=\langle t_1,\ldots,t_{m+2} \rangle \rtimes \langle t_0\rangle$, and because $\sigma_1=t_0$ it follows that the set of variables (\ref{vars}) is indeed in correspondence with the complete conjugacy class of $\sigma_1$.
The motivation for using restricted group $\langle t_1,\ldots,t_{m+2}\rangle$ instead of the full group $\langle t_0,t_1,\ldots,t_{m+2}\rangle$ (as in \cite{ib}) is because it does not lose generality, and is convenient for our subsequent analysis connecting with the Painlev\'e systems.


The manifestation of the 5-simplex consistency here is the well-posedness of the following initial-value-problem.
\begin{prop}\label{latgeom}
Impose a 5-simplex consistent rational triplet-pair system (Definitions \ref{tps} and \ref{5sc}) on all triplet-pairs of the arrangement of Definition \ref{Gtp}.
Then all remaining variables are composed rational functions of the unconstrained subset
\begin{equation}
w(\sigma_1^\omega),\ldots,w(\sigma_m^\omega),w(\sigma_1),\ldots,w(\sigma_m).\label{fvars}
\end{equation}
\end{prop}
In the case $m=2$ this is the statement of rationality of the triplet-pair system, whilst in the case $m=3$ it coincides with statement of the 5-simplex consistency.\\
{\bf Remark.}
Combined with the group of symmetries of the arrangement, this initial-value-problem allows to derive constructively the preceding birational group, the main points of the derivation are as follows.
The natural action of generators $\sigma_1,\ldots,\sigma_m,\omega$ permute variables (\ref{fvars}) with other variables in the lattice, and according to Proposition \ref{latgeom} the other variables can then be expressed as rational functions of variables (\ref{fvars}).
The identification of variables listed in (\ref{fvars}) with those listed in (\ref{Evars}) then allows to identify the rational action induced by the generators with the previously defined rational actions (\ref{sigmadef}), (\ref{omegadef}) in the case $n=m$.
This closes a circle, obtaining the birational group from the initial-value-problem on the lattice, and demonstrates the equivalence of the birational-group and lattice-system approaches.

\subsection{Restriction to the quad-graph}\label{RTQG}
The lattice of Definition \ref{Gtp} is a generalisation to the usual quad-graph domain of the Yang-Baxter maps \cite{ABSf}.
Specifically, the hypercube of dimension $m$.
This subsection exhibits the quad-graph as restriction from the more general one.

It corresponds to restriction to the subgroup 
\begin{equation}
\label{hdef}
H:= \langle t_3,t_4,\ldots,t_{m+2}\rangle 
\end{equation}
of the group defined in Proposition \ref{vmg}, which is specified here in terms of a subset of the group elements introduced in (\ref{iso}).
Subgroup $H$ defines a sub-arrangement as follows.
\begin{definition}\label{mcube}
The triplet-pair arrangement of the $m$-cube is the subset of variables (\ref{vars}) given by
\begin{equation}\label{subvars}
w(\sigma_1^{g\omega}),\ w(\sigma_1^g), \quad g\in H,
\end{equation}
arranged into the subset of the triplet-pairs (\ref{gc}) given by
\begin{equation}
w(\sigma_1^{g}),w(\sigma_2^{g\omega}),w(\sigma_2^{g\sigma_1\omega}), \quad
w(\sigma_2^{g}),w(\sigma_1^{g\omega}),w(\sigma_1^{g\sigma_2\omega}), \quad
g\in H.\label{subgc}
\end{equation}
\end{definition}
Again this is consistent because $\sigma_2=\sigma_1^{t_4}$, and $t_4\in H$.

The $m$-cube lattice geometry is made completely clear by introducing alternative coordinates as follows:
\begin{equation}
\sigma_I := \prod_{i\in I}\sigma_i, \quad I\subseteq \{1,\ldots,m\},
\end{equation}
in which ordering of the composition is not important because $\sigma_i$ and $\sigma_j$ commute.
\begin{prop}
The triplet-pair arrangement of the $m$-cube (Definition \ref{mcube}) can be written differently as variables
\begin{equation}\label{cubeidents}
w(\sigma_i^{\sigma_I\omega}),\ w(\sigma_i), \quad i\in\{1,\ldots,m\}, \ I\subseteq\{1,\ldots,m\},
\end{equation}
arranged into triplet-pars
\begin{equation}\label{ctps}
w(\sigma_i),w(\sigma_j^{\sigma_I\omega}),w(\sigma_j^{\sigma_{I\oplus \{i\}}\omega}), \quad
w(\sigma_j),w(\sigma_i^{\sigma_I\omega}),w(\sigma_i^{\sigma_{I\oplus \{j\}}\omega}),
\end{equation}
where $i,j\in\{1,\ldots,m\}$, $i\neq j$, and $I\subseteq\{1,\ldots,m\}$. 
Furthermore, the action of the generators of $H$ (\ref{hdef}) on variables (\ref{cubeidents}) reduce to action on $i$ and $I$ as follows
\begin{equation}\label{DN}
\begin{split}
&t_3t_4 = \{I \mapsto I \oplus \{1,2\}\},\\
&t_{i+3} = \{ I \mapsto I|_{i\leftrightarrow i+1}, i\leftrightarrow i+1\}, \quad i\in\{1,\ldots,m-1\}.
\end{split}
\end{equation}
\end{prop}
Here $\oplus$ denotes the symmetric difference of sets, $I\oplus J = (I\cup J)\setminus (I\cap J)$.

The subsets $I\subseteq\{1,\ldots,m\}$ are in natural correspondence with vertices of an $m$-cube whose edges connect vertices $I$ and $I\oplus \{i\}$.
The index $i\in\{1,\ldots,m\}$ is in correspondence with the $i^{th}$ {\it characteristic} of the $m$-cube, which is the $(m-1)$-dimensional section that bisects all edges connecting vertex $I$ with vertex $I\oplus \{i\}$, $I\subseteq\{1,\ldots,m\}$.
Thus variables $w(\sigma_1),\ldots,w(\sigma_m)$ are associated with characteristics and are considered global parameters of the restricted system, which are known in this context as lattice parameters. 
The other kind of variable $w(\sigma_i^{\sigma_I\omega})$ is associated with the hypercube edge connecting vertex $I$ with vertex $I\oplus \{i\}$.
The triplet-pairs (\ref{ctps}) are associated with quads of the $m$-cube, they involve variables on four edges around a quad and parameters on characteristics which intersect on that quad.
The initial data (\ref{fvars}) consists of variables on all edges attached to the origin vertex $I=\{\}$, and the set of lattice parameters.

Note that in (\ref{DN}) the action of $t_3t_4$ has been given, instead of giving $t_3$ directly.
The given action is clear because $t_3t_4=\sigma_1\sigma_2$ according to (\ref{iso}), and it is easy to see that on variables (\ref{cubeidents}) we have the action $\sigma_j = \{I \mapsto I \oplus \{j\}\}$, which shows the origin of $\sigma_1,\ldots,\sigma_m$, as basic reflections of the $m$-cube.
The Yang-Baxter property is encoded by the commutativity $\sigma_i\sigma_j=\sigma_j\sigma_i$ corresponding to the first of relations (\ref{rels2}).

Also note that the action of $H$ on variables (\ref{cubeidents}) preserves the parity of $|I|$, but every edge of the $m$-cube is attached to a unique vertex with $|I|$ even, so although Definition \ref{mcube} leads always to the evenness of $|I|$, this does not lose generality.\\
{\bf Remark.}
It has been mentioned in Section \ref{INTRO} that in this setting, the Yang-Baxter maps are manifestations of the discrete KdV equation.
The explicit connection is due to Papageorgio, Tongas and Veselov \cite{PTV}, who established relation with the integrable quad-equation introduced much earlier by Nijhoff, Quispel and Capel (NQC) in \cite{NQC}.
The NQC equation is a generalisation of the potential-KdV superposition formula of Wahlquist and Estabrook \cite{WE}, but is equivalent to it if we allow third-order non-local transformations \cite{NAH,James2}.

\section{Linearisation of the actions $t_1,\ldots,t_{m+2}$}\label{LTI}
In Section \ref{ABG}, a birational representation of the Coxeter group $\langle t_1,\ldots,t_{m+2}\rangle$ on the variables $x_1,\ldots,x_n,y_1,\ldots,y_n$, (valid for $1<m\le n$) was recalled.
In this section, we give a linearisation of these actions.
The relation with the birational actions arising in the theory of the elliptic Painlev\'e equation\cite{sc} is also established.

\subsection{The case of $F_I$}\label{TCF1}
In this subsection, we consider the case of the primary triplet-pair system (\ref{F1a}).
To linearize the corresponding birational actions $t_1,\ldots,t_{m+2}$, we introduce variables
$h_1, h_2, k_1, k_2,\ldots, k_{n+3}$, and define a parametrization of variables $x_1,\ldots,x_n$, $y_1,\ldots,y_n$ as
\begin{equation}\label{xy-param}
\begin{split}
x_i=\dfrac{c_{1,3}c_{2,i+3}}{c_{2,3}c_{1,i+3}}, &\qquad
c_{i,j}=[k_i-k_j][h_1-k_i-k_j], \\
y_i=\dfrac{d_{1,3}d_{2,i+3}}{d_{2,3}d_{1,i+3}},&\qquad
 d_{i,j}=[k_i-k_j][h_2-k_i-k_j],
\end{split}
\end{equation}
where $[u]$ is the Weierstrass sigma function
$[u]=\sigma(u)$ (elliptic case),
$[u]={\rm sinh}(u)$ (trigonometric case) or  $[u]=u$ (rational case).

\begin{thm}\label{thm:linear}
Under the substitution (\ref{xy-param}),
the actions of $t_1,\ldots,t_{m+2}$ (defined in Section \ref{ABG}) on variables
$(x_i,y_i)_{i=1,\ldots,n}$
are compatible with the following linear actions on
$h_1, h_2, k_1, k_2, \ldots, k_{n+3}$:
\begin{equation}\label{t-linear-action}
\begin{split}
t_1 & = \{h_1 \leftrightarrow h_2\},\\
t_2 & =  \{h_i \mapsto h_i+\beta, k_{i\leq 4} \mapsto k_i+\beta\},\\
& \qquad \beta=h_1+h_2-k_1-k_2-k_3-k_4,\\
t_3 & = \{h_1 \mapsto h_1+h_2-k_4 - k_5, k_4 \mapsto h_2-k_5, k_5 \mapsto h_2-k_4\},\\
t_{j} & = \{k_{j+1} \leftrightarrow k_j\}, \quad (j=4,\ldots,m+2).
\end{split}
\end{equation}
\end{thm}

\noindent
{\it Proof}. 
We prove the Theorem by using
the following simpler actions $s_0, \ldots, s_{m+3}$.
They act linearly on
$h_1, h_2, k_1, k_2, \ldots, k_{n+3}$ and birationally on $(x_i,y_i)_{i=1,\ldots, n}$ as
\begin{equation}\label{s-action}
\begin{split}
s_0 & =\{k_{1}\leftrightarrow k_{2}, x_{i}\to \dfrac{1}{x_{i}}, y_{i}\to \dfrac{1}{y_{i}}\},\\
s_1 & =\{h_1 \leftrightarrow h_2, x_i\leftrightarrow y_i\},\\
s_2 & =\{h_{1}\to h_{1}+h_{2}-k_{1}-k_{2}, k_{1}\to h_{2}-k_{2}, k_{2}\to h_{2}-k_{1},\\
& \qquad x_{i}\to \dfrac{x_{i}}{y_{i}}, y_{i}\to \dfrac{1}{y_{i}}\},\\
s_3 & =\{k_{2}\leftrightarrow k_{3}, x_{i}\to 1-x_{i}, y_{i}\to 1-y_{i}\},\\
s_4 & =\{k_{3}\leftrightarrow k_{4}, x_{1}\to \dfrac{1}{x_{1}}, x_{i>1}\to \dfrac{x_{i}}{x_{1}}, y_{1}\to \dfrac{1}{y_{1}}, y_{i>1}\to \dfrac{y_{i}}{y_{1}}\},\\
s_i & =\{k_{i-1}\leftrightarrow k_{i}, x_{i-4}\leftrightarrow x_{i-3}, y_{i-4}\leftrightarrow y_{i-3}\}, \quad (i=5,\ldots,m+3).
\end{split}
\end{equation}
It is easy to check that $s_0, s_1, \ldots, s_{m+3}$ generate the Coxeter group associated with
the following Dynkin diagram:
\[
\begin{array}{cccccccccccccc}
&&&& s_0\\
&&&& \vert\\
s_1&-&s_2&-&s_3&-&s_4&-&\cdots&-&s_{m+3}.\\
\end{array}
\]
Moreover, these actions are compatible with the substitutions (\ref{xy-param}).
The compatibilities for actions $s_i, (i\neq 3)$ are satisfied without any condition for the function $[u]$.
The compatibility for $s_3$ reduces to the equations
\begin{equation}
c_{12}c_{34}-c_{13}c_{24}+c_{14}c_{23}=0,\quad
d_{12}d_{34}-d_{13}d_{24}+d_{14}d_{23}=0,
\end{equation}
which are equivalent to the Riemann relation for $[u]$.

Due to the compatibility of the actions $s_i$ with the substitution (\ref{xy-param}), the compatibility of the actions $t_i$ follows if
they are realised as some compositions of $s_i$'s.
Such compositions are easily found by looking at the linear actions
(\ref{t-linear-action}) which are realised by
\begin{equation}
\begin{split}
t_1 & = s_1,\\
t_2 & = s_2 \mu s_2,\\
t_3 & = \nu s_2 \nu,\\
t_i & = s_{i+1}, \quad (i=4,\ldots,m+2)\\
\mu & = s_1s_3s_0s_4s_3s_2s_3s_0s_4s_3s_1,\\
\nu & = s_3s_4s_3s_0s_5s_3s_4s_3.
\end{split}\label{F1iso}
\end{equation}
Our final task is to confirm that these compositions reproduce the correct actions as $t_i$ also on variables $(x_i,y_i)_{i=1,\ldots,n}$.
To do this, we compute the explicit actions of $\mu, \nu$, which are given as
\begin{equation}
\begin{split}
\mu & = \{k_{3}\to h_{1}-k_{4}, k_{4}\to h_{1}-k_{3}, h_{2}\to h_{1}+h_{2}-k_{3}-k_{4},\\
&\qquad x_{1}\to \dfrac{1}{x_{1}}, x_{i>1}\to \dfrac{x_{i}}{x_{1}}, y_{1}\to \dfrac{y_{1}}{x_{1}}, y_{i>1}\to \tilde{y}_i\},\\
\nu & = \{k_1\leftrightarrow k_4, k_2\leftrightarrow k_5, \\
&\qquad x_{1}\to \dfrac{1-x_{1}}{1-x_{2}}, x_{2}\to \dfrac{(1-x_{1}) x_{2}}{x_{1} (1-x_{2})}, x_{i>2}\to \dfrac{(1-x_{1}) (x_{2}-x_{i})}{(1-x_{2}) (x_{1}-x_{i})},\\
&\qquad y_{1}\to \dfrac{1-y_{1}}{1-y_{2}}, y_{2}\to \dfrac{(1-y_{1}) y_{2}}{y_{1} (1-y_{2})}, y_{i>2}\to \dfrac{(1-y_{1}) (y_{2}-y_{i})}{(1-y_{2}) (y_{1}-y_{i})}\},
\end{split}
\end{equation}
where $\tilde{y}_{i>1}=\mu(y_i)$ is determined from 
$\dfrac{\tilde{y}_i-\mu(y_1)}{\tilde{y}_i-1}=\dfrac{y_i-y_1}{y_i-1}\dfrac{x_i-1}{x_i-x_1}.$
Then one can verify the following relations 
\begin{equation}
\begin{array}{l}
t_2(x_1)=y_1, \quad t_2(y_1)=x_1,\\
\dfrac{R(x_i,\{t_2(x_i),x_1,y_i\})}{R(y_1,\{t_2(x_i),x_1,y_i\})}=\dfrac{R(x_i,C)}{R(y_1,C)}, \\
\dfrac{R(y_i,\{t_2(y_i),y_1,x_i\})}{R(x_1,\{t_2(y_i),y_1,x_i\})}=\dfrac{R(y_i,C)}{R(x_1,C)}, 
\quad (i=2,3,\ldots,n)
\end{array}
\end{equation}
and
\begin{equation}
\begin{array}{l}
\dfrac{R(x_1,\{t_3(x_1),x_2,y_1\})}{R(y_2,\{t_3(x_1),x_2,y_1\})}=\dfrac{R(x_1,C)}{R(y_2,C)},\\
\dfrac{R(x_2,\{t_3(x_1),x_1,y_2\})}{R(y_1,\{t_3(x_1),x_1,y_2\})}=\dfrac{R(x_2,C)}{R(y_1,C)},\\
\dfrac{t_3(x_i)-t_3(x_1)}{t_3(x_i)-t_3(x_2)}=\dfrac{y_i-y_2}{y_i-y_1}\dfrac{x_i-x_1}{x_i-x_2},\\
t_3(y_1)=y_2, \quad t_3(y_2)=y_1,\quad t_3(y_i)=y_i, \quad (i=3,\ldots,n)
\end{array}
\end{equation}
where $R(x,\{c_1,c_2,c_3\})=(x-c_1)(x-c_2)(x-c_3)$ and $C=\{0,1,\infty\}$.
These are the characteristic properties of the birational actions of $t_2, t_3$, hence the Theorem \ref{thm:linear} is proved.[]

\subsection{Relation to the elliptic Painlev\'e equation}

In previous subsection the Coxeter group actions $t_1,\ldots, t_{m+2}$ are recovered through the actions $s_0,\ldots,s_{m+3}$ given in (\ref{s-action}). The latter actions and their linearisation have been essentially known in the theory of the discrete Painlev\'e equation. 
We will give a short summary about these materials in a convention suitable for our description.

The elliptic difference Painlev\'e equation, the master equation among the second order discrete Painlev\'e equations, was discovered and constructed by Sakai \cite{sc} in $\P^2$ form. The fundamental device in the construction is the birational representation of the affine Weyl group $\tilde{E}_8$. The elliptic Painlev\'e equation is  obtained as the translation part of the affine Weyl group.

In our formulation the $\tilde{E}_8$ action in $\P^1 \times \P^1$ form is more suitable than $\P^2$.
But both formulations are essentially the same and we will describe them in parallel way. Let 
\begin{equation}
U=\left[\begin{array}{ccccc} 
u_1&u_2&\cdots&u_{n}\\
v_1&v_2&\cdots&v_{n}\\
\end{array}\right]\in {\rm Mat}_{2,n}(\P^1),
\end{equation}
be a configuration of $n$ points $p_i=(u_i,v_i)$ on $\P^1\times \P^1$ and consider their natural equivalent class $\mathcal{M}_{n}$ given by
\begin{equation}
\mathcal{M}_{n}={\rm PGL}(2,\C)^{\otimes 2}\setminus\ {\rm Mat}_{2,n}(\P^1),
\end{equation}
where the M\"obius transformation ${\rm PGL}(2,\C)$ acts on each low of $U$ diagonally.
We will consider in open chart whose coordinates are given by the canonical form
\begin{equation}
U_{\rm can}=\left[\begin{array}{ccccccc} 
\infty&0&1&f_4&\cdots&f_{n}\\
\infty&0&1&g_4&\cdots&g_{n}
\end{array}\right],
\end{equation}
where
\begin{equation}
f_i=\frac{u_{2i}u_{13}}{u_{1i}u_{23}}, \quad
g_i=\frac{v_{2i}v_{13}}{v_{1i}v_{23}},
\end{equation}
and $u_{ij}=u_i-u_j, v_{ij}=v_i-v_j$.

Alternatively, let
\begin{equation}
X=\left[\begin{array}{ccccc} 
x'_1&x'_2&\cdots&x'_{n+1}\\
y'_1&y'_2&\cdots&y'_{n+1}\\
z'_1&z'_2&\cdots&z'_{n+1}
\end{array}\right]\in {\rm Mat}_{3,n+1}(\C),
\end{equation}
be a configuration of $n+1$ points $p'_i=(x'_i:y'_i:z'_i)$ on $\P^2$ and consider their natural equivalent class $\mathcal{M}'_{n+1}$ given by
\begin{equation}
\mathcal{M}'_{n+1}={\rm PGL}(3,\C)\setminus\ {\rm Mat}_{3,n+1}(\C) /{(\C^{*})^{n+1}},
\end{equation}
We have an open chart whose coordinates are given by the canonical form
\begin{equation}
X_{\rm can}=\left[\begin{array}{ccccccc} 
1&0&0&1&f'_5&\cdots&f'_{n+1}\\
0&1&0&1&g'_5&\cdots&g'_{n+1}\\
0&0&1&1&1&\cdots&1
\end{array}\right],
\end{equation}
where
\begin{equation}
f'_{i}=\frac{\mu_{23i}\mu_{124}}{\mu_{12i}\mu_{234}}, \quad
g'_{i}=\frac{\mu_{13i}\mu_{124}}{\mu_{12i}\mu_{134}}, 
\end{equation}
and $\mu_{ijk}$ is a minor determinant of $X$ taking $i$, $j$ and $k$-th columns.

There exist a natural birational map between the configuration spaces $\mathcal{M}_n$ and $\mathcal{M}'_{n+1}$ which is simply given by identification of the coordinates $(f_i,g_i)=(f'_{i+1}, g'_{i+1})$ and hence $p_i=p'_{i+1}$ ($i=2,\ldots,n$). This birational map gives the isomorphism between $X_n=($$n$ points blowing up of $\P^1\times \P^1$)
and $X'_{n+1}=$ ($n+1$ points blowing up of $\P^2$) for $n \geq 1$. We will identify these spaces $X_n=X'_{n+1}$:
\begin{equation}
\begin{array}{ccccc}
&&X'_2=X_1\\
&\hskip-5mm p'_1, p'_2&& \quad p_1&\\
&\swarrow&&\searrow&\\
\P^2&&&&\P^1\times \P^1.
\end{array}
\end{equation}

On the configuration space $\mathcal{M}_n \sim \mathcal{M}'_{n+1}$, we 
consider birational maps $s_0, s_1, \ldots, s_{n}$ coming from permutations of points as follows:
\begin{center}
\begin{tabular}{|c||c|c|}
\hline
action & meaning on $\P^2$ & meaning on $\P^1\times \P^1$\\
\hline\hline
$s_0$ & a standard Cremona action           & $p_1\leftrightarrow p_2$\\
\hline
$s_1$ & $p'_1\leftrightarrow p'_2$ & permutation of two $\P^1$'s\\
\hline
$s_2$ & $p'_2\leftrightarrow p'_3$ & a quadratic transformation\\
\hline
$s_{i\geq 3}$ & $p'_i\leftrightarrow p'_{i+1}$ & $p_{i-1}\leftrightarrow p_i$\\
\hline
\end{tabular}
\end{center}
The explicit actions are easily computed as follows
\begin{equation}\label{s-on-fg}
\begin{split}
s_0&=\{ f_i \rightarrow \frac{1}{f_i}, g_i \rightarrow \frac{1}{g_i}\},\\
s_1&=\{ f_i \leftrightarrow g_i \},\\
s_2&=\{ f_i \rightarrow \frac{f_i}{g_i}, g_i \rightarrow \frac{1}{g_i}\},\\
s_3&=\{ f_i \rightarrow 1-f_i, g_i \rightarrow 1-g_i\},\\
s_4&=\{ f_4 \rightarrow \frac{1}{f_4}, f_{i>4} \rightarrow \frac{f_i}{f_4},
 g_4 \rightarrow \frac{1}{g_4}, g_{i>4} \rightarrow \frac{g_i}{g_4}\},\\
s_{i>4}&=\{ f_{i-1}\leftrightarrow f_i, g_{i-1}\leftrightarrow g_i\},
\end{split}
\end{equation}
which gives a birational representation of the Coxeter group
\begin{equation}
\begin{array}{ccccccccccccc}
&&&& s_0\\
&&&& \vert\\
s_1&-&s_2&-&s_3&-&s_4&-&\cdots&-&s_{n}.\\
\end{array}
\end{equation}
The actions (\ref{s-on-fg}) explain the birational action of $s_i$ (\ref{s-action}) under the renumbering of variables $f_i=x_{i-3}, g_i=y_{i-3}$.
Similarly, the linear actions of $s_i$ (\ref{s-action}) on parameters $h_1, h_2, k_1, k_2, \ldots$ have also the geometric origin given in terms of the Picard lattice of $X_n=X'_{n+1}$.
We have canonical basis of the Picard lattices as follows
\begin{equation}
\begin{array}{l}
{\rm Pic}(X_n)=\Z H_1\oplus \Z H_2 \oplus \Z E_1 \oplus \cdots \oplus \Z E_n,\\
{\rm Pic}(X'_{n+1})=\Z \mathcal{E}_0 \oplus \Z {\mathcal E}_1 \oplus \cdots \Z {\mathcal E}_{n+1},
\end{array}
\end{equation}
where the non-vanishing intersection pairings are given by
\begin{equation}
\begin{array}{c}
(H_1,H_2)=(H_2,H_1)=1,\quad  (E_i,E_i)=-1,\\
({\mathcal E}_0, {\mathcal E}_0)=1,\quad  ({\mathcal E}_i, {\mathcal E}_i)=-1. \quad (i\geq 1)
\end{array}
\end{equation}
The isomorphism ${\rm Pic}(X_n)={\rm Pic}(X'_{n+1})$ is given by
\begin{equation}
\begin{array}c
{\mathcal E}_0=H_1+H_2-E_1-E_2, \quad
{\mathcal E}_1=H_1-E_1, \quad
{\mathcal E}_2=H_2-E_1, \quad
{\mathcal E}_{i\geq 3}=E_{i-1}, \\
H_1={\mathcal E}_0-{\mathcal E}_2, \quad
H_2={\mathcal E}_0-{\mathcal E}_1, \quad
E_1={\mathcal E}_0-{\mathcal E}_1-{\mathcal E}_2, \quad
E_{i\geq 2}={\mathcal E}_{i+1}.
\end{array}
\end{equation}
The action on the Picard lattice induced by $s_i$ is then given by the simple reflection
\begin{equation}
s_{i}(\beta)=\beta +(\alpha_i, \beta)\alpha_i
\end{equation}
with the corresponding root $\alpha_i$ given by
\begin{center}
\begin{tabular}{|c||c|c|}
\hline
root & ${\rm Pic}(X'_{n+1})$ & ${\rm Pic}(X_n)$\\
\hline\hline
$\alpha_0$ & ${\mathcal E}_0-{\mathcal E}_1-{\mathcal E}_2-{\mathcal E}_3$          & $E_1-E_2$\\
\hline
$\alpha_1$ & ${\mathcal E}_1-{\mathcal  E}_2$ & $H_1-H_2$\\
\hline
$\alpha_2$ & ${\mathcal E}_2-{\mathcal E}_3$ & $H_2-E_1-E_2$\\
\hline
$\alpha_{i\geq 3}$ & ${\mathcal E}_i-{\mathcal E}_{i+1}$ & $E_{i-1}-E_i$\\
\hline
\end{tabular}
\end{center}
Then by introducing the complex parameters as
$\epsilon_i, h_i, k_i$ corresponding  to ${\mathcal E}_i , H_i, E_i$,
we have the following linear actions $s_i$ on them
\begin{equation}
\begin{array}{l}
s_0=\{k_1 \leftrightarrow k_2\},\\
s_1=\{h_1\leftrightarrow h_2\},\\
s_2=\{h_1 \rightarrow h_1+h_2-k_1-k_2, k_1 \rightarrow h_2-k_2,
k_2 \rightarrow h_2-k_1\},\\
s_{i\geq 3}=\{ k_{i-1} \leftrightarrow k_{i} \},
\end{array}
\end{equation}
or equivalently
\begin{equation}
\begin{array}{l}
s_0=\{\epsilon_0 \rightarrow 2\epsilon_0-\epsilon_1-\epsilon_2-\epsilon_3, \
\epsilon_i\rightarrow\epsilon_0-\epsilon_j-\epsilon_k, \ \{i,j,k\}=\{1,2,3\} \},\\
s_{i\geq 1}=\{ \epsilon_{i-1} \leftrightarrow \epsilon_{i} \}.
\end{array}
\end{equation}

The parametrization (\ref{xy-param}) which is written in the notation here as
\begin{equation}
\begin{split}
x_{i-3}&=f_i=
\frac{[k_2-k_i][h_1-k_2-k_i]}{[k_1-k_i][h_1-k_1-k_i]}
\frac{[k_1-k_3][h_1-k_2-k_i]}{[k_2-k_i][h_1-k_1-k_3]},\\
y_{i-3}&=g_i=
\frac{[k_2-k_i][h_2-k_2-k_i]}{[k_1-k_i][h_2-k_1-k_i]}
\frac{[k_1-k_3][h_2-k_2-k_i]}{[k_2-k_i][h_2-k_1-k_3]},
\end{split}
\end{equation}
is obtained from
\begin{equation}
(u_i,v_i)=\left(
\frac{[k_2-k_i][h_1-k_2-k_i]}{[k_1-k_i][h_1-k_1-k_i]},
\frac{[k_2-k_i][h_2-k_2-k_i]}{[k_1-k_i][h_2-k_1-k_i]}\right),
\end{equation}
or equivalently
\begin{equation}
(x'_i:y'_i:z'_i)=\left(
\frac{[\epsilon_0-\epsilon_2-\epsilon_3-\epsilon_i]}{[\epsilon_1-\epsilon_i]}:
\frac{[\epsilon_0-\epsilon_3-\epsilon_1-\epsilon_i]}{[\epsilon_2-\epsilon_i]}:
\frac{[\epsilon_0-\epsilon_1-\epsilon_2-\epsilon_i]}{[\epsilon_3-\epsilon_i]}\right).
\end{equation}
The linearisation in the last $\P^2$ form was known (see \cite{kmnoy}  for example).

\noindent
{\bf Remark.} 
The actions (\ref{s-on-fg}) gives the affine Weyl group of type $\tilde{E}_8$ for $n=8$, In the theory of Painlev\'e equations, however, we consider the case $n=9$ (not $8$). The reason is that the last point $p_9$ plays special role (the dependent variable of the discrete Painlev\'e equation), and we omit the last generator $s_9$ to get the affine Weyl group of type $\tilde E_8$. This kind of formulation of the eliptic Painlev\'e equation where the dependent variable is treated like an additional blowing up point was first considered in \cite{WE10} in $\P^2$ formulation.

Accordingly, in the Painlev\'e context, the variables $(f_i,g_i)_{i=4,\ldots,8}$ can be  parametrized by $h_1,h_2,k_1,\ldots,k_8$ as above, but not for the last point $(f_9,g_9)$. Since the last point $p_9$ is generically not on the bidegree $(2,2)$ curve passing through the eight points $p_1, \ldots, p_8$.

\noindent
{\bf Remark.} Recall that the actions $t_1,t_3,t_4,\ldots$ give a birational representation of the Coxeter group 
\begin{equation}
\begin{array}{ccccccccccccc}
&&&& t_4\\
&&&& \vert\\
t_1&-&t_3&-&t_5&-&t_6&-&\cdots.\\
\end{array}
\end{equation}
This representation can be considered as a version of the actions (\ref{s-action}) or (\ref{s-on-fg}) where the normalization conditions on the first three points $p_1=(\infty,\infty), p_2=(0,0), p_3=(1,1)$ are relaxed. 
In fact, in eq. (\ref{F1iso}), $t_3$ and $s_2$ was intertwined by the transformation $\nu$ which exchanges the parameters $k_1,k_2$ (related to the points $p_1=(\infty,\infty), p_2=(0,0)$) with $k_4,k_5$ (related to the points $p_4=(x_1,y_1), p_5=(x_2,y_2)$).
Then the actions $t_3=\nu s_2 \nu$ and $t_1=s_1$, $t_4=s_5, t_5=s_6, \ldots$ do not touch upon the variables $k_1,k_2,k_3$ (or $p_1,p_2,p_3$) and act only on the un-normalized points
$p_4, p_5, \ldots$. In this way, we obtained the Coxeter group action where the normalization constraint for the three points are removed. We note that such a normalization free representation was only available by using some elliptic functions (see \cite{msy}) 
and purely algebraic form without elliptic function parametrisation was not known before.

\subsection{Degenerations to $F_{II}$ and $F_{IV}$}
In this subsection, we will derive the linearisation of the birational actions $t_1,\ldots,t_{m+2}$ associated with the triplet-pair systems (\ref{F2a}) and (\ref{F4a}) as suitable degenerations from those already presented for (\ref{F1a}).

Take as starting point the key substitution (\ref{xy-param}):
\begin{multline}\label{F1sol}
x_i=F(k_{i+3},h_1),\ y_i=F(k_{i+3},h_2),\\ F(u,h)=\frac{[u-k_2][h-k_2-u][k_3-k_1][h-k_1-k_3]}{[u-k_1][h-k_1-u][k_3-k_2][h-k_2-k_3]}.
\end{multline}
The limit $k_2\rightarrow k_1$ implies that $F(u,h)\rightarrow 1+(k_2-k_1)\hat{F}(u,h)+O((k_2-k_1)^2)$ where
\begin{equation}\label{F2sol}
\hat{F}(u,h)=\zeta(k_3-k_1)+\zeta(h-k_1-k_3)-\zeta(u-k_1)-\zeta(h-k_1-u).
\end{equation}
The further limit $k_1\rightarrow k_3$ implies that $ \hat{F}(u,h)\rightarrow 1/(k_3-k_1)+{\doublehat{F}(u,h)}+O((k_3-k_1))$ where
\begin{multline}\label{F4sol}
{\doublehat{F}}(u,h)=\frac{1}{2}\frac{\wp'(u-k_3)-\wp'(h-k_3-u)}{\wp(u-k_3)-\wp(h-k_3-u)} \\ = \zeta(h-2k_3)-\zeta(u-k_3)-\zeta(h-k_3-u).
\end{multline}
The participating functions are written for the elliptic case, but more generally they are defined in terms of $[u]$ by the equations
\begin{equation}
\zeta(u) = \frac{[u]'}{[u]}, \quad \wp(u) = \frac{([u]')^2-[u][u]''}{[u]^2}.
\end{equation}
So, limiting cases are included as follows:
\begin{equation}
\begin{array}{cclcl}
\hline
{\rm ell}&\rightarrow&{\rm trig}&\rightarrow&{\rm rat}\\
\hline
\sigma(u) &\rightarrow& {\mathrm{sinh}}(u) &\rightarrow& u \\
\zeta(u) &\rightarrow& {\mathrm{cosh}}(u)/{\mathrm{sinh}}(u) &\rightarrow& 1/u\\
\wp(u) &\rightarrow& 1/{\mathrm{sinh^2}}(u) &\rightarrow& 1/u^2\\
\hline
\end{array}
\end{equation}

The degeneration chain for the underlying triplet-pair systems is very straightforward.
The transition $(\ref{F1a})\rightarrow (\ref{F2a})$ can be made by the substitution $x_i\rightarrow 1+\epsilon x_i,y_i\rightarrow 1+\epsilon y_i$, and the transition $(\ref{F2a})\rightarrow (\ref{F4a})$ can be made by substitution $x_i\rightarrow 1/\epsilon+x_i,y_i\rightarrow 1/\epsilon+y_i$.
These are consistent with the above limits on $F$ which have been made via coalescence of the parameters $k_1$, $k_2$ and $k_3$ ($\epsilon=k_2-k_1$ for $(\ref{F1a})\rightarrow (\ref{F2a})$ and $\epsilon=k_1-k_3$ for $(\ref{F2a})\rightarrow (\ref{F4a})$).
As a result, we have the following:
\begin{prop}\label{subs2}
Under the substitutions
$x_i=\hat{F}(k_{i+3},h_1)$, 
$y_i=\hat{F}(k_{i+3},h_2)$
and 
${x}_i={\doublehat{F}}(k_{i+3},h_1)$, 
${y}_i={\doublehat{F}}(k_{i+3},h_2)$
respectively, the action of mappings $t_1,\ldots,t_{m+2}$ (\ref{iso}) constructed from respective underlying rational triplet-pair systems (\ref{F2a}) and (\ref{F4a}), are consistent with the linear actions (\ref{t-linear-action}) in the admissible special cases $k_2=k_1$ and $k_1=k_2=k_3$ respectively.
\end{prop}

To develop the geometric understanding of the degenerate cases, we can complement the remark given at the end of Section \ref{TCF1} as follows.
\begin{prop}\label{nlt}
Through associations
\begin{equation}\label{cr}
f_{i+3}=\frac{(x_{i+3}-x_2)(x_3-x_1)}{(x_{i+3}-x_1)(x_3-x_2)},\qquad
g_{i+3}=\frac{(y_{i+3}-y_2)(y_3-y_1)}{(y_{i+3}-y_1)(y_3-y_2)},
\end{equation}
the rational action of \[t_4,t_1,t_3,t_5,t_6,\ldots,t_{m+2}\] on variables $x_1,\ldots,x_{n}$,$y_1,\ldots,y_n$ constructed in Section \ref{ABG} from any of (\ref{F1a}), (\ref{F2a}) or (\ref{F4a}), induce, respectively, the actions \[s_0,s_1,s_2,s_3,s_4,\ldots,s_{m}\] on variables $f_4,\ldots,f_{n}$, $g_4,\ldots,g_{n}$, given in (\ref{s-on-fg}).
\end{prop}
{\it Proof.} The normalizing associations (\ref{cr}) quotient out any M\"obius change of the original variables, and it is not difficult to see that the aforementioned limiting procedures suffer a similar fate, so the proposition follows directly from the comment at the end of Section \ref{TCF1}.[]\\
{\bf Remark.}
Proposition \ref{nlt} shows that the distinction between the birational groups (omitting generators $t_0$ and $t_2$) associated with the three triplet-pair systems (\ref{F1a}), (\ref{F2a}) and (\ref{F4a}) is lost within the geometrically admissible normalization that puts $(x_1,y_1)$, $(x_2,y_2)$, $(x_3,y_3)$ at $(\infty,\infty)$, $(0,0)$, $(1,1)$.
From the setting of the associated lattice system (Section \ref{AL}) this can be seen as existence of non-local transformations, modulo which the three systems are equivalent.
On reflection, this is not unexpected: it extends the similar fact which is already known to be true on the restricted quad-graph domain (cf. the remark at the end of Section \ref{RTQG}).

\subsection{As a solution of the lattice initial-value-problem}
In this subsection we comment on aspects of the established connection from the view of the lattice geometry.

The established linearisation of birational actions $t_1,\ldots,t_{m+2}$ corresponds to integration of the following particular initial value problem (cf. Proposition \ref{latgeom}):
\begin{equation}\label{livp}
w(\sigma_i^\omega) = F(k_{i+3},h_1),\quad
w(\sigma_i) = F(k_{i+3},h_2),\quad i\in\{1,\ldots,m\},
\end{equation}
where $h_1,h_2,k_1,\ldots,k_{m+3}$ are freely chosen.
The participating function $F$ was given in (\ref{F1sol}), and is appropriate for the triplet-pair system (\ref{F1a}), but should be replaced with $\hat{F}$ (\ref{F2sol}) or $\doublehat{F}$ (\ref{F4sol}) for the corresponding triplet-pair systems (\ref{F2a}) and (\ref{F4a}).
The solution to initial value problem (\ref{livp}) takes the form
\begin{equation} \label{lsol}
w(\sigma_1^g) = F(g(k_{4}),g(h_2)), \quad g\in\langle t_1,\ldots,t_{m+2}\rangle.
\end{equation}
Here the action of $g$ on the left permutes elements of the conjugacy class of $\sigma_1=t_0$, to which the lattice variables are assigned.
The action on the right is the linear action (\ref{t-linear-action}).

The Painlev\'e sub-lattice corresponds to the case $m\ge 10$ and $n=m+1$, and the subset of variables
\begin{equation}\label{psl}
w(\sigma_{m+1}^g),\  g \in \langle t_1,t_3,t_4,\ldots,t_{10}\rangle,
\end{equation}
or any set conjugate to this one.
The remaining variables (\ref{lsol}) are considered as non-autonomous parameters.
The initial data $x_{m+1}=w(\sigma_{m+1}^\omega)$ and $y_{m+1}=w(\sigma_{m+1})$ are freely chosen and complement (\ref{livp}).

One can consider the natural extension of the Painlev\'e sub-lattice (\ref{psl}) to the remaining variables
\begin{equation}\label{mgs}
w(\sigma_{m+1}^g),\  g \in \langle t_1,t_2,t_3,\ldots,t_{m+2}\rangle.
\end{equation}
Such Coxeter group representation is interesting.
We don't comment on the the relation to integrable dynamics for the general group, however we note that (according to Proposition \ref{latgeom}) the extension from (\ref{psl}) to (\ref{mgs}) is unique.

We also note the restriction to the different sub-lattice corresponding to the quad-graph domain $w(\sigma_{m+1}^g),\ g\in\langle t_3,t_4,\ldots,t_{m+2}\rangle$.
The natural hypercube coordinates (cf. Section \ref{RTQG}) allow to express the solution (\ref{lsol}) more explicitly on the hypercube sub-lattice, it looks as follows:
\begin{equation}\label{hcsol}
\begin{split}
w(\sigma_i^{\sigma_I\omega}) & = F(k_{i+3},h_1+|I|h_2/2-{\textstyle \sum_{j\in I}}k_{j+3}), \quad i\not\in I,\\
& = F(h_2-k_{i+3},h_1+|I|h_2/2-{\textstyle \sum_{j\in I}} k_{j+3}), \quad i\in I,\\
\end{split}
\end{equation}
where $I\subseteq \{1,\ldots,m\}$ with $|I|$ even.
Notice that $k_4,\ldots,k_{m+3}$ correspond to lattice parameters in this setting.
The restriction of variables (\ref{mgs}) to this sub-lattice, that is, the subset of variables $w(\sigma_{m+1}^g): g\in\langle t_3,t_4,\ldots,t_{m+2}\rangle$, are in correspondence with edges in one singled-out direction of the $(m+1)$-cube, and thus vertices of the $m$-cube.
Integration on this domain corresponds to applying the B\"acklund transformation to (\ref{hcsol}), which is treated as an elliptic background, or seed solution, of the associated Yang-Baxter maps.
The independent equation satisfied by the vertex variables themselves is the corresponding multi-quadratic quad-equation \cite{AtkNie} which is naturally understood as the consequence of the Yang-Baxter maps \cite{ib}.

\vskip5mm

\noindent
{\bf Acknowledgments.} 
JA acknowledges support from the Australian Research Council, Discovery Grant DP 110104151.
YY is supprted by JSPS KAKENHI Grant Number 26287018.

\vskip5mm

\bibliographystyle{unsrt}
\bibliography{references}

\end{document}